\documentclass[11pt,a4paper,oneside, twocolumn]{article}
\usepackage[bindingoffset=0in, left=1in, right=1in, top=1in, bottom=1in]{geometry}
\usepackage{abstract}

\newcommand{\beginsupplement}{%
        \setcounter{table}{0}
        \renewcommand{\thetable}{S\arabic{table}}%
        \setcounter{figure}{0}
        \renewcommand{\thefigure}{S\arabic{figure}}%
     }

\usepackage{changepage}
\usepackage{subcaption}
\usepackage[utf8]{inputenc}
\usepackage{comment}
\usepackage{amsmath}
\usepackage{amssymb}
\usepackage{color}
\usepackage{graphicx}
\usepackage{hyperref}
\usepackage{fancyhdr}
\usepackage[switch]{lineno}
\usepackage{upgreek}
\usepackage{authblk}
\usepackage{enumitem}
\usepackage[dvipsnames]{xcolor}
\setenumerate{itemsep=0mm}

\usepackage{titlesec}
\titleformat*{\section}{\normalsize\bfseries\centering}

\title{\vspace{-10mm} \bf \normalsize First Measurement of Correlated Charge Noise in Superconducting Qubits at an Underground Facility\vspace{-2mm}}

\author[2,1]{G.~Bratrud} 
\author[3,1]{S.~Lewis}
\author[4,1]{K.~Anyang}
\author[2]{A.~Colón~Cesaní}
\author[5,6,7]{T. Dyson}
\author[1,5,6,7,8]{H.~Magoon}
\author[2]{D.~Sabhari}
\author[9,1]{G.~Spahn}
\author[1]{G.~Wagner}
\author[2]{R.~Gualtieri}
\author[1,6,7]{N.A.~Kurinsky}
\author[1]{R.~Linehan}
\author[9]{R.~McDermott}
\author[1]{S.~Sussman}
\author[1]{D.J.~Temples}
\author[1]{S.~Uemura}
\author[10]{C.~Bathurst}
\author[1]{G.~Cancelo}
\author[2]{R.~Chen}
\author[1]{A.~Chou}
\author[4,1]{I.~Hernandez}
\author[1]{M.~Hollister}
\author[1]{L.~Hsu}
\author[1]{C.~James}
\author[2]{K.~Kennard}
\author[4,1]{R.~Khatiwada}
\author[1]{P.~Lukens}
\author[2]{V.~Novati\thanks{Presently at LPSC, Centre National de la Recherche Scientifique, Université Grenoble Alpes, Grenoble, France}}
\author[2]{N.~Raha}
\author[2]{S.~Ray}
\author[2]{R.~Ren\thanks{Presently at Department of Physics, University of Toronto, Toronto, ON, Canada}}
\author[2]{A.~Rodriguez}
\author[2]{B.~Schmidt\thanks{Presently at IRFU, Alternative Energies and Atomic Energy Commission, Université Paris-Saclay, France}}
\author[1,6,7]{K.~Stifter}
\author[4]{J.~Yu}
\author[1,2]{D.~Baxter}
\author[2,1]{E.~Figueroa-Feliciano}
\author[1]{D.~Bowring \thanks{Corresponding author: dbowring\@fnal.gov}}

\affil[1]{Fermi National Accelerator Laboratory, Batavia, IL 60510, USA}
\affil[2]{Department of Physics \(\&\) Astronomy, Northwestern University, Evanston, IL 60208, USA}
\affil[3]{Department of Physics and Astronomy, Wellesley
College, Wellesley, MA 02481, USA}
\affil[4]{Department of Physics, Illinois Institute of Technology, Chicago, IL 60616, USA}
\affil[5]{Department of Physics, Stanford University, Stanford, CA 94305, USA}
\affil[6]{Kavli Institute for Particle Astrophysics and Cosmology, Stanford University, Stanford, CA 94305, USA}
\affil[7]{SLAC National Accelerator Laboratory, Menlo Park, CA 94025, USA}
\affil[8]{Department of Physics \& Astronomy, Tufts University, Medford, MA 02155, USA}
\affil[9]{Department of Physics, University of Wisconsin-Madison, Madison, WI 53706, USA}
\affil[10]{Department of Physics, University of Florida, Gainesville, FL 32611, USA}

\date{\vspace{-20pt} \small (Dated: \today)}

\begin{document}

\twocolumn[
\begin{@twocolumnfalse}
\maketitle
\begin{abstract}
\begin{adjustwidth}{32pt}{32pt}
\vspace{-25pt}

We measure space- and time-correlated charge jumps on a four-qubit device, operating 107 meters below the Earth's surface in a low-radiation, cryogenic
facility designed for the characterization of low-threshold particle detectors. 
The rock overburden of this facility reduces the cosmic ray muon flux by over $99\%$ compared to laboratories at sea level. Combined with $4\pi$ coverage of a movable lead shield, this facility enables quantifiable control over the flux of ionizing radiation on the qubit device.
Long-time-series charge tomography measurements on these weakly charge-sensitive qubits capture
discontinuous jumps in the induced charge on the qubit islands, corresponding to
the interaction of ionizing radiation with the qubit substrate. 
The rate 
of these charge jumps scales with the flux of ionizing radiation on the
qubit package, as characterized by a series of independent measurements
on another energy-resolving detector operating simultaneously in the same cryostat with the 
qubits.
Using lead shielding, we achieve a minimum charge jump rate of $0.19 ^{+0.04}_{-0.03}$~mHz, almost an order of magnitude lower than that measured in surface tests, but a factor of roughly seven higher than expected based on reduction of ambient gammas alone. 
We operate four qubits for over 22 consecutive hours with zero correlated charge jumps at length scales above three millimeters.
\end{adjustwidth}
\end{abstract}
\end{@twocolumnfalse}]

\saythanks

A growing body of evidence indicates that ionizing radiation impacts the coherence of superconducting qubits. Correlation has been measured between the flux of ionizing radiation and the qubit energy relaxation rate ($1/T_1$) \cite{vepsalainen20, gusenkova22}. Ionizing events in a chip substrate cause simultaneous errors in multi-qubit processors \cite{mcewen22}. Moreover, the presence of correlated qubit errors, and the rate at which they occur in unshielded laboratories, can interfere with the efficacy of error-correcting surface codes \cite{google23, Martinis:2020fxb}. 
Researchers have observed charge and parity errors, correlated in time and space, caused by environmental gammas and cosmic rays interacting with the qubit substrate \cite{christensen19, wilen21, harrington24}.
Ionizing radiation has also been shown to ``scramble'' the spectrum of two-level system populations in superconducting qubits~\cite{Thorbeck:2022yzs}.


Here we describe a continuation of the work performed in  Ref.~\cite{wilen21}, using the same four-qubit chip. 
The charge environment in the qubit substrate responds to ionizing events through multiple physics channels (including e-h pair production, phonon diffusion, and charge trapping) and on time scales ranging from nanoseconds to hours or even days \cite{ramanathan20,devissier11,mcewen22}. 
Accordingly, the qubits in this work are mildly
charge-sensitive transmons ($E_J/E_C=24$) that operate as electrometers, each with a sensing area for electric fields in the substrate of hundreds of square microns. 
The qubit chip was relocated from the Earth's surface at Madison, WI to the Northwestern EXperimental Underground Site (NEXUS) at Fermilab in Batavia, IL. 
Previously, correlated jumps in offset charge, associated with gamma ray and cosmic ray impacts, were observed in this qubit array.  
Underground, over 99$\%$ of cosmic ray muons are shielded by the overburden, creating an environment in which the qubit response to gamma radiation can be studied in isolation. 
We vary the flux of gamma rays incident on the qubit chip and measure the rate and magnitude of ensuing charge burst events via Ramsey tomography. 
Other energy-resolving detectors operating simultaneously in NEXUS are used to calibrate the flux and spectrum of ionizing radiation. 
(This study is in some sense complimentary to that of Ref.~\cite{harrington24}, in which scintillation detectors provide coincidence information between cosmic ray muon events and qubit errors.)


%
%
NEXUS is a low-background test stand for cryogenic detector calibration at Fermilab \cite{temples24, chen24}. 
The facility is located 107~m underground in the Neutrinos at the Main Injector (NuMI/MINOS) beam line access tunnel, with a rock and concrete overburden corresponding to 225~meters water equivalent~\cite{Adamson:2015}. 
With this overburden, the muon flux from cosmic rays is a factor of 200 lower than at a surface facility: approximately 7~muons/cm$^2$/day, with a negligible hadronic shower rate~\cite{Garrison:2014nla}. 
The Madison qubit package was installed in a Cryoconcept HEXA-DRY pulse-tube dilution refrigerator (DR) with passive vibration isolation and a modular lead shield that attenuates environmental radiation with 4$\pi$ coverage. 
The DR and all experimental electronics operate in a class 10,000 clean room to minimize sources of particulate contamination and radioactivity from dust. Additional facility and experimental hardware details are discussed in the Supplemental Material.

%
For different datasets during this run period, we vary the flux of gammas incident on the qubit chip by opening and closing the lead shield, and we measure the rate of discontinuous changes in offset charge (``charge jumps''), both
on individual qubits and the correlated event rate across pairs of qubits. 
The work presented here consists of two datasets corresponding to the lead shield being open (ambient gamma flux) and closed (minimal gamma flux). 
The integrated measurement times in the ``Shield Open" (SO) and ``Shield Closed" (SC) configurations are 23.949 and 22.075~hours respectively. 

\begin{figure}[t]
    \begin{subfigure}[b]{\textwidth}
        \includegraphics[width=0.45\textwidth]{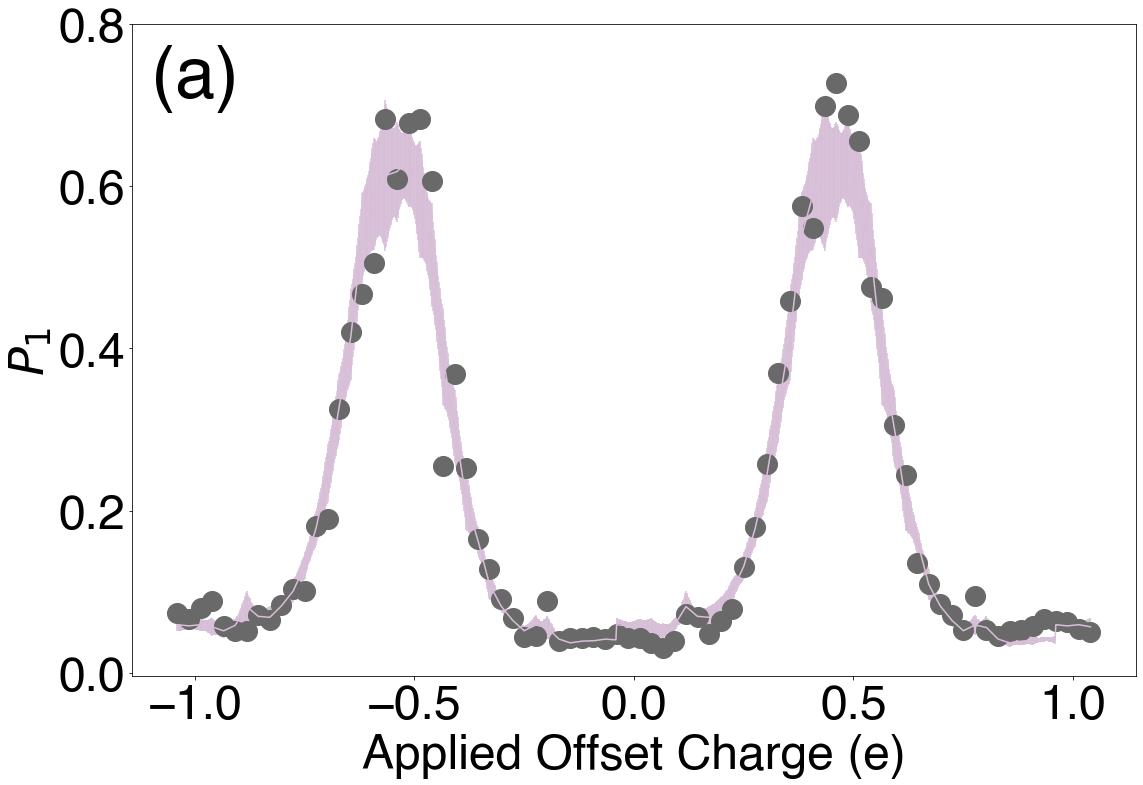}
        \label{fig:ramsey_analytic}
    \end{subfigure}
    \begin{subfigure}[b]{\textwidth}
        \includegraphics[width = 0.45\textwidth]{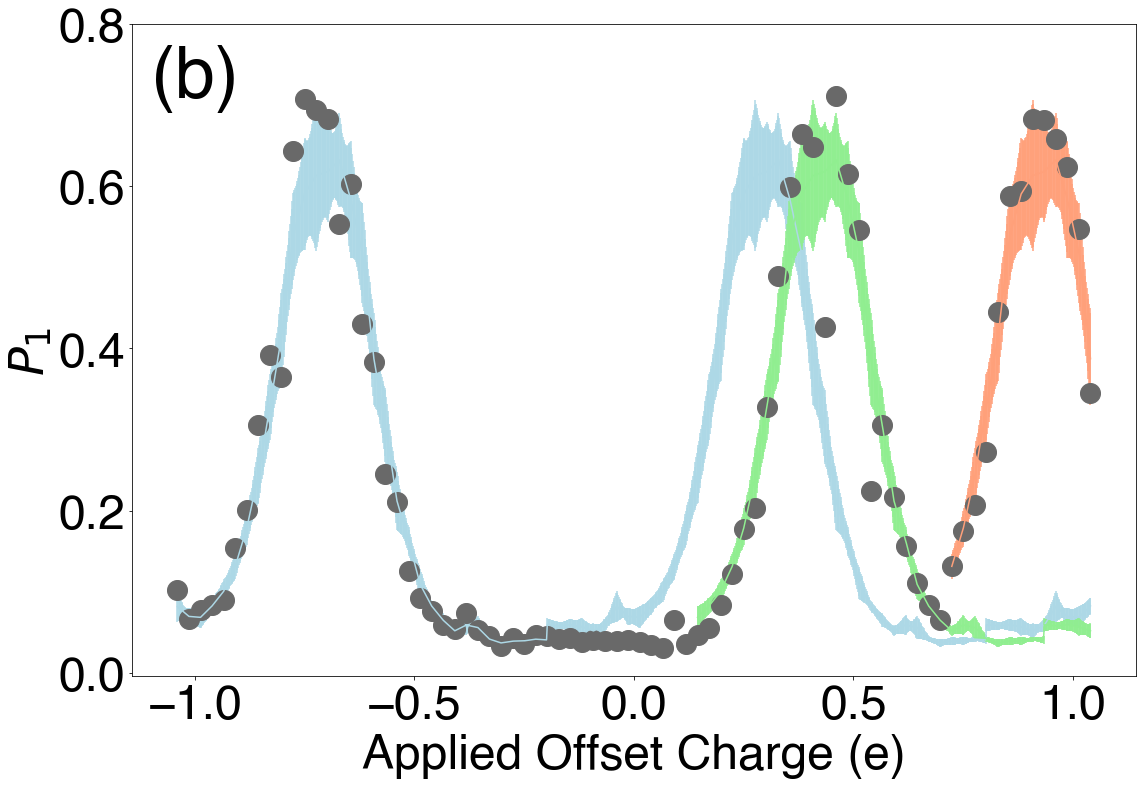}
        \label{fig:jumpscan}
    \end{subfigure}
    \caption{(a) Example of a single charge tomography scan from one qubit. Each gray point is the average of 200
    measurements. The purple band shows a fit to the template, with the width representing standard error across template samples.
    (b) A second tomography scan consisting of two charge jumps, with the shaded lines corresponding to the new best template fit after each jump. The green (orange) shaded template corresponds to a charge jump with \text{$\Delta q = 0.13e~(0.50e) \pm 0.03e$} relative to the previous template.}\label{fig:ramsey}
\end{figure}

\begin{table*}[t!]
\renewcommand{\arraystretch}{1.3}
\centering
\begin{tabular}{r c c c}\hline
& Shield Open & Shield Closed & Units \\\hline
Livetime & 23.949 & 22.075 & hours \\
Q1 Rate & $0.42^{+0.09}_{-0.08}$ & $0.20^{+0.07}_{-0.05}$ & mHz \\
Q2 Rate & $0.60^{+0.11}_{-0.09}$ & $0.19^{+0.07}_{-0.05}$ & mHz \\
Q3 Rate & $0.52^{+0.10}_{-0.08}$ & $0.19^{+0.07}_{-0.05}$ & mHz \\
Q4 Rate & $0.51^{+0.11}_{-0.09}$ & $0.16^{+0.07}_{-0.05}$ & mHz \\\hline
Average Rate & $0.51^{+0.05}_{-0.04}$ & $0.19^{+0.04}_{-0.03}$ & mHz \\
Corrected $\gamma$ Rate & $0.34^{+0.07}_{-0.06}$ & $0.02^{+0.06}_{-0.05}$ & mHz\\
Calculated Excess Rate & \multicolumn{2}{c}{$0.17^{+0.04}_{-0.03}$} & mHz \\\hline
\end{tabular}
\caption{Efficiency-corrected rates of charge jumps with magnitude $0.1e \le |\Delta q| \le 0.5e$. Only statistical errors are shown; systematic errors are an order of magnitude smaller. Given the apparent non-dependence on external gamma flux for the Shield Closed (SC) data, we subtract this from the Shield Open (SO) data to find the rate associated only with gamma impacts, as well as the excess jump rate not associated with external gammas. ``Livetime'' here refers to the total time interval over which data was continuously collected. 
}\label{tab:sing_rates}
\end{table*}

\begin{table*}[t!]
\renewcommand{\arraystretch}{1.3}
\centering
\begin{tabular}{ r c c c c c c c } \hline
   & Q1-Q2 & Q3-Q4 & Q1-Q3 & Q1-Q4 & Q2-Q3 & Q2-Q4 & Units \\ \hline
  Separation & 640 & 340 & 3195 & 3330 & 3180 & 3240 & $\upmu$m \\
Shield Open & $\rm 0.27 ^{+0.09}_{-0.07}$ & $\rm 0.29 ^{+0.09}_{-0.07} $ & $\rm 0.03 ^{+0.04}_{-0.02} $ & $\rm 0.08 ^{+0.06}_{-0.04} $ & $\rm 0.05 ^{+0.05}_{-0.03} $ & $\rm 0.08 ^{+0.06}_{-0.04} $ & mHz \\
Shield Closed &  $\rm 0.10 ^{+0.07}_{-0.04} $ & $\rm 0.04 ^{+0.05}_{-0.03} $ & $\rm < 0.03 $ & $\rm < 0.04 $ & $\rm < 0.03 $ & $\rm < 0.04 $ & mHz  \\ \hline
\end{tabular}
\caption{Efficiency-corrected rates (mHz) of correlated charge jumps with magnitude $0.1e \le |\Delta q| \le 0.5e$ in each qubit pair. Statistical errors are provided; systematic errors are an order of magnitude smaller. The separation distances of each qubit pair are provided for reference; see Figure \ref{fig:cornerplot}. We measure zero correlated charge jumps for qubits separated by over 3~mm over 22 hours of continuous data taking.}\label{tab:cor_rates}
\end{table*}

A Ramsey sequence ($X/2-Idle-X/2$) is
applied to each qubit. During the idle period $t_{\mathrm{idle}}$ of this sequence, the state vector phase $\phi$ evolves as a function 
of the offset charge $n_g$ present on the qubit island, 
\begin{equation}\label{eq:phase}
    \phi(n_g)=\Delta f_{01}t_{\rm idle}\cos(2\pi n_g),
\end{equation}
for charge dispersion $\Delta f_{01}$. Here, $n_g$ is the sum of the charge applied via the bias
lines and the intrinsic offset charge.
This pulse sequence maps gate charge (modulo 1$e$) onto the excited state probability $P_1$ of the qubit. A discontinuous 
change in $n_g$, as from an induced electric field following a charge burst event, causes a discontinuity in $P_1$, as shown in Figure \ref{fig:ramsey}. These discontinuities are recorded as charge jumps with a magnitude $\Delta q$. 
Scanning the charge bias on each qubit across a range of voltages allows for a calibration of $n_g$ values and an extraction of $P_1$ that is less sensitive to charge noise.
See Supplemental Material for further details.


In principle, $n_g$ can be analytically determined from the excited state population $P_1$~\cite{christensen19}.
However, when multiple burst  events occur in the qubit substrate 
during a single tomographic scan, and/or when other incoherent
noise is present in the system, fitting against this 
functional form is not an efficient method of detecting 
charge jumps. 
Instead, we perform a rolling $\chi^2$ minimization to fit each tomographic scan against a template averaged from approximately 20 jump-free scans, with the ``phase'' of $P_1$ floating on $n_g\in(-0.5e , +0.5e)$. 
Examples of a jumpless scan and a scan with two detected jumps are shown in Figure~\ref{fig:ramsey}. 

\begin{figure*}[t!]
\centering
\includegraphics[width=\linewidth]{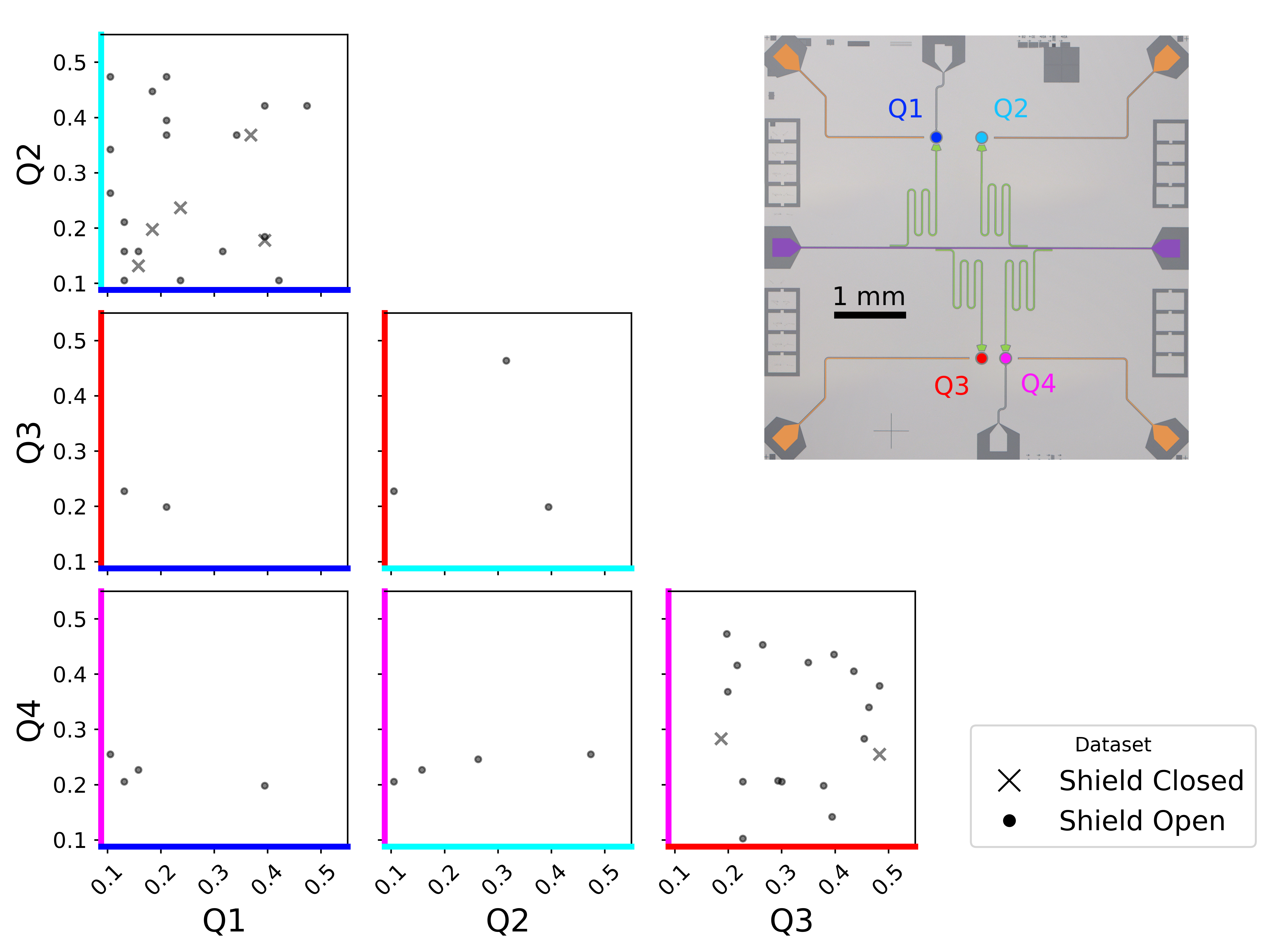}
\caption{Correlated jump magnitudes $0.1e \le |\Delta q| \le 0.5e$ for all qubit pairs, for SC (x's) and SO (dots), in units of electric charge, $e$.  The raw data retains sign information, but only magnitude $|\Delta q|$ is presented here for visual clarity. Non-correlated events are omitted, also for visual clarity. (Inset) A micgrograph of the qubit chip, annotated with false colors to match the plot axes.}
\label{fig:cornerplot}
\end{figure*}

To quantify the efficiency of this method, the equivalent of 
400 hours of Ramsey tomography scans were simulated for each qubit,
and convolved with a Gaussian noise spectrum according to the measured noise in each qubit. Charge jumps
are simulated by injecting $P_1$ discontinuities into this dataset
at varying intervals and with varying sizes. In all four qubits, this method finds an 
efficiency of 
$>$70\% for identifying jumps of magnitude \text{$0.1e \le |\Delta q| \le 0.5e$}, with larger values of $|\Delta q|$ aliased down to the masurement interval.
See Supplemental Materials for details.
The above procedure is independently run on the scans from each of the four 
qubits, with charge burst rates extracted for
single qubits and for all qubit pairs. The resulting rates above a charge 
jump threshold of $|\Delta q| \ge 0.1 e$ for each qubit are displayed in Table~\ref{tab:sing_rates}. The corresponding time-correlated rates across qubit pairs are displayed in Table~\ref{tab:cor_rates} and the relative jump sizes $|\Delta q|$ of these pairs are shown in Figure~\ref{fig:cornerplot}. A pair of charge jumps is considered time-correlated if they occur within 44~s of each other. 

We compare the rate of qubit charge jumps against the flux of gamma rays present during the collection of SO and SC datasets. These fluxes were measured by a $\rm Li_2MoO_4$ (LMO) crystal instrumented with a Transition Edge Sensor as a thermistor~\cite{LMO} and located 18.7~cm from the qubit chip in the DR. We thus perform a direct measurement of the ratio 
between SC and SO gamma fluxes.
Based on the LMO data, we determine that the gamma flux in the SO case should be a factor of  $20 \pm 1$ larger than in the SC case, for gamma energies above 150~keV.

The qubit payload was also exposed to $^{133}$Ba and $^{137}$Cs sources. The lead shield was closed during the collection of this data, with the sources positioned inside the lead shield but outside the cryostat. The $^{133}$Ba data, used to calibrate our LMO detector energy scale, had an integrated gamma rate close to the SO data (see Figure~\ref{fig:LMO} in the Supplemental Material). Operational issues at the underground facility limited our $^{133}$Ba source exposure time to six hours, such that no statistically meaningful comparison can be made between this data set and the SO and SC data sets discussed in this paper. However, the measured jump rates at these low statistics are similar to the SO jump rates, as expected from a gamma-dominated scenario. Similarly, a $^{137}$Cs source was installed to study the qubit response to higher gamma fluxes. This $^{137}$Cs source saturated the LMO detector, making spectrum calibrations impossible in that configuration. The $^{133}$Ba and $^{137}$Cs data sets are therefore not included in this analysis. Future work will focus in more depth on the use of external gamma sources to vary the flux and spectrum of ionizing radiation on this qubit package.

The charge jump rates, averaged across all four qubits, that we measure for the SO and SC data are 0.51 and 0.19~mHz respectively, as shown in Table~\ref{tab:sing_rates}. 
Therefore, closing the shield reduces the rate of qubit charge bursts by only a factor of 2.7, over seven times less than expected based on the factor of 20 reduction in gamma flux measured in the LMO detector. 
A possible explanation for this discrepancy is that in our lowest-background SC configuration, we are sensitive to an excess source of charge bursts that is not dominated by the external gamma flux. 
Potential sources of radiation inside the fridge that might significantly impact the qubit package but not the LMO detector warrant follow up study and assay~\cite{Loer:2024xrl,Cardani:2022blq}. 
The expected muon flux through the qubit (for both configurations) is $\sim$0.08~mHz/cm$^2$ -- too low to explain the observed rates. 
The NuMI muon neutrino beam was not active during data collection for this work. Neutrino interactions in matter \cite{adamson12} therefore contributed no additional muon flux during this study. 

We infer that the ambient gamma flux does not dominantly contribute to our SC dataset. We therefore subtract the SC rates from the SO data to obtain a reduced burst rate in the SO data, induced by ambient gammas, of $0.34^{+0.07}_{-0.06}$~mHz. 
In subtracting these data to determine the gamma-induced component, we account for an unknown population of bursts by assuming a constant excess rate of jumps present in both the SO and SC data. 
Based on the LMO data (see Supplemental Material), we estimate our ambient gamma flux in the SO configuration to be approximately five times lower than that measured using a NaI detector in Ref.~\cite{wilen21}.  
If we assume, as Ref.~\cite{wilen21} does, that the surface burst rate of 1.35~mHz was gamma-dominated, 
our ambient rate estimate of $0.34$~mHz is proportional with that expected reduction in flux. 

The distances between each qubit pair are different, per Table \ref{tab:cor_rates} and Figure \ref{fig:cornerplot}. The smallest separation between qubit pairs (qubits 3 and 4) is 340~$\upmu$m and the largest separation between pairs (qubits 1 and 4) is 3330~$\upmu$m. This variable separation between qubit pairs enables some inferences about correlated noise rates. 
First, it is technically possible that separate burst events could create conditions that mimic correlated charge jumps arising from a single charge burst. 
In the SO data, we measure a correlated charge jump rate in nearby qubit pairs of 0.27~mHz for qubits 1 and 2, and 0.29~mHz for qubits 3 and 4. 
Given the low single-qubit jump rates in Table \ref{tab:sing_rates}, the rate at which this stochastic coincidence could occur is orders of magnitude lower than the measured correlated jump rate. 
Next, this correlated rate is approximately half of the single-qubit rate (to within statistical uncertainty), which is consistent with that observed in Ref.~\cite{wilen21} when dominated by gamma flux. 
In the SO data, the correlated jump rate across distant qubit pairs is too low to make any statements on burst origin with any statistical significance. 
The same statistical limitation is true for nearby qubit pairs in the SC data. 
Finally, we are able to, for the first time, eliminate correlated charge noise in charge-sensitive qubits separated by over 3~mm, on timescales nearing one day.

We therefore present the first results from a charge-sensitive qubit chip operated in an underground environment. 
We observe a reduction in charge burst events commensurate with the reduction in ambient gamma flux relative to Ref.~\cite{wilen21}. 
Furthermore, in our low-background SC dataset, we observe an excess of charge bursts that appears inconsistent with both the expected muon rates and the ambient gamma flux.
The next steps will be to investigate the origin of these excess charge bursts; candidates for this origin include trapped charge in the substrate that relaxes on long timescales, secondaries from cosmogenic interactions with the DR materials, and the presence of an anomalous radiation source very close to the qubit chip in the DR. 
Further study of these events in low-background environments is required to better understand the mechanisms of charge bursts and other errors and their impact on qubit performance.
Despite this unexplained excess, our lowest background dataset is free of correlated charge jumps at length scales above 3~mm during approximately one day of continuous operation. 
In addition to the implications for fault-tolerant quantum computing, an 
understanding of the effects of ionizing radiation on the performance of
superconducting qubits is critical to the development of these devices for use as particle detectors for fundamental physics \cite{dixit21, Linehan:2024niv}.

The authors would like to thank Steve Hahn and Cindy Joe for their management of the MINOS underground facility. 
The authors would like to thank Adam Anderson for providing the QICK board used for these measurements. The TWPA 
was provided by MIT-Lincoln Laboratory and IARPA.

This manuscript has been authored by Fermi Research Alliance, LLC under Contract No. DE-AC02-07CH11359 with the U.S. Department of Energy, Office of Science, Office of High Energy Physics. 
This work was supported by the U.S. Department of Energy, Office of Science, National Quantum Information Science Research Centers, Quantum Science Center, and the U.S. Department of Energy, Office of Science, High-Energy Physics Program Office.  
This work was supported in part by the U.S. Department of Energy, Office of Science, Office of Workforce Development for Teachers and Scientists (WDTS) under the Science Undergraduate Laboratory Internships Program (SULI). 
This material is based upon work supported by the National Science Foundation Graduate Research Fellowship Program under Grant No. DGE-2234667. Any opinions, findings, and conclusions or recommendations expressed in this material are those of the authors and do not necessarily reflect the views of the National Science Foundation. 

GB led the analysis of the result included in this paper, with help from ACC and GW. 
SL led the RF upgrade of the NEXUS fridge and installation of the qubit payload, with help from DBo, TD, HM, GS, NK, VN, BS, and JY. 
GB, SL, KA, RL, and HM contributed to the data-taking scripts used to make this measurement. 
DS led the analysis of the LMO detector data, with help from RG, RC, VN, and BS. 
RG, RL, SS, DJT, and KS provided crucial input on qubit operation. 
IA, SR, and JY provided input through parallel analyses outside of the scope of this paper. 
SU and GC provided support with the QICK board and firmware used to take this data. 
VN, BS, and DJT led the operation and maintenance of the NEXUS facility, with help from DBa, DBo, CB, RC, EFF, RG, MH, CJ, KK, PL, NR, RR, AR, and LH. 
DBa, DBo, EFF, NK, and RM provided guidance in the scoping and presentation of this work.
DBa, DBo, GC, AC, EFF, LH, and RK provided leadership of the local quantum group at Fermilab. 
RM provided the qubit chip used in this measurement. 
DBo coordinated and led this effort through his DOE Early Career Award. 
All authors provided feedback and thoughtful discussions throughout the development of this work.


\bibliographystyle{ieeetr}
\bibliography{references}


\clearpage
\section*{\large Supplemental Material}


\beginsupplement
\renewcommand\thesection{\ifcase\value{section}?\or A\or B\or C\or D\or E\or F\else G\fi}

\section{Description of experimental apparatus}

\begin{table}[b]
\centering
\begin{tabular}{ l | c c c c }
 & Q1 & Q2 & Q3 & Q4 \\ \hline
\rule{0pt}{2.5ex}Resonator (GHz) & 6.18 & 5.82 & 6.07 & 5.95 \\
\rule{0pt}{2.5ex}$f_{01}$ (GHz) & 4.83 & 4.71 & 4.53 & 4.69 \\
\rule{0pt}{2.5ex}$\Delta f_{01}$ (MHz) & 2.6 & 3.1 & 3.9 & 3.4 \\
\end{tabular}
\caption{Measured frequencies of devices used. Resonator frequency denotes the frequency of the readout mode, $f_{01}$ is the qubit transition frequency, and $\Delta f_{01}$ is the frequency dispersion.}\label{tab:freq}
\end{table}

The 6.25 x 6.25 mm$^2$ sample chip incorporates four weakly charge-sensitive circular transmon qubits, as shown in Figure~\ref{fig:cornerplot}. Each qubit consists of a circular superconducting Nb island set within a circular hole in the superconducting Nb groundplane (respective radii $r_i = 70$~$\upmu$m and $r_o = 90.5$~$\upmu$m), with one Al/AlO$_x$/Al Josephson junction bridging the gap. For a uniform electric field, the sensing area of a single qubit is $\pi\epsilon r_ir_o$, with relative permittivity $\epsilon$. Each qubit has a ratio of Josephson energy to single-electron charging energy $E_J/E_C = 24$ and is capacitively coupled to an offset charge control line (Fig. \ref{fig:cornerplot}, yellow) as well as a readout resonator (Fig. \ref{fig:cornerplot}, green) for dispersive measurement through a shared feedline (Fig. \ref{fig:cornerplot}, purple). The qubits are split into two pairs, one on each side of the central feedline, with center-to-center intra-pair separations of 340~$\upmu$m and 640~$\upmu$m, and inter-pair separation of around 3~mm. Additional qubit parameters are available in Tables \ref{tab:cor_rates} and \ref{tab:freq}, and in Ref.~\cite{wilen21}.

\begin{figure}[t]
    \centering
    \includegraphics[width=\linewidth]{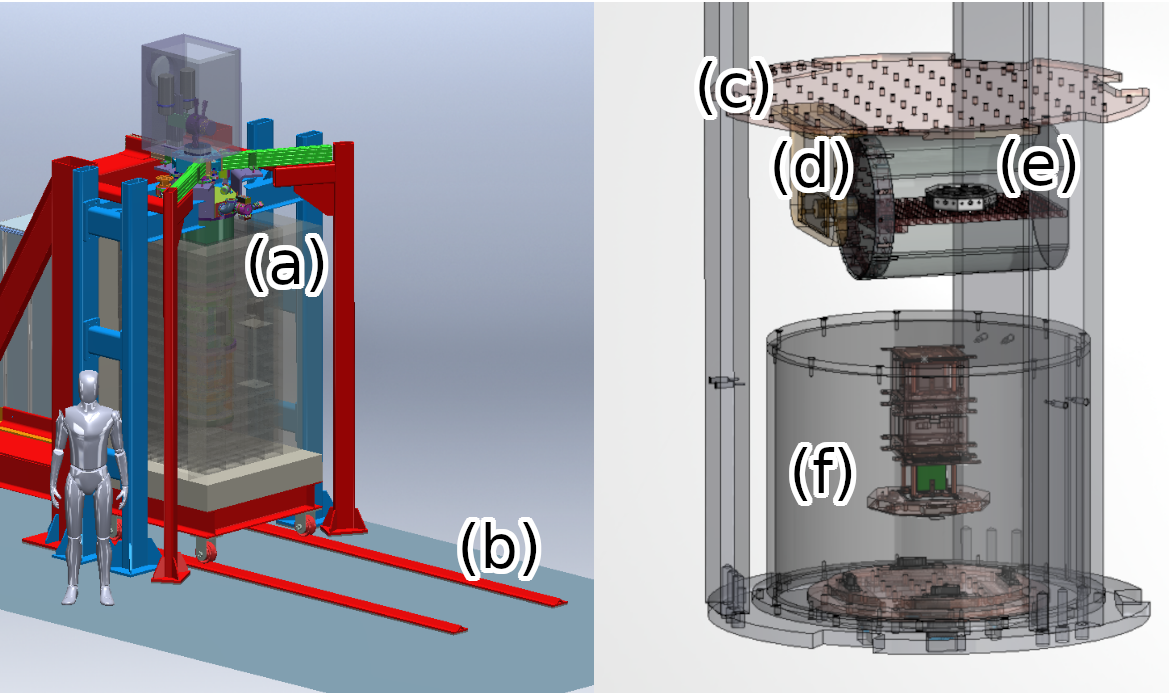}
    \caption{Schematic of NEXUS DR and shield assembly (left) and expermintal payloads in the DR (right). (a) The movable lead shield covers three sides of the DR. It rests on a wheeled platform. (The fourth side of the lead shield is permanently installed on the DR frame and platform, and is not visible in this schematic. (b) The shield platform rolls on a set of rails, permanently installed in the floor of the clean room. (c) NEXUS DR 10~mK plate. (d) A4K magnetic shield for qubit package. (e) Qubit package. (f) LMO detector package. The straight-line distance between the center of the qubit chip and the center of the LMO crystal is 18.7~cm. Cabling, instrumentation, thermalization, and other miscellaneous hardware and instrumentation are not shown.}\label{fig:nexus}
\end{figure}

The NEXUS DR, shown in Figure~\ref{fig:nexus}, is enclosed in a three-part lead shield: (1) a four-inch-thick inner lead plug above the payload, thermalized at $\sim$1~K; (2) a four-inch-thick stationary wall; and (3) a cart-mounted, movable lead shield with a nominal thickness of four inches. 
Taken together, these three shield components provide full, $4\pi$ coverage of experimental payloads against ambient gamma radiation. 
The qubit chip is installed inside of a 1-mm thick A4K hermetic can from Amuneal, thermalized to the DR mixing chamber plate. 
The DR itself is further equipped with a 1-mm thick A4K magnetic shield from Amuneal, thermalized at 4~K to reduce 
the magnetic field in the payload region. 
Finally, a external Metglass blanket surrounds the 300-K shell. 
The DR base temperature was stabilized at 10.5~mK 
during the collecation of all data presented here.

\section{Electronics chain}

\begin{figure*}[t!]
\centering
\includegraphics[width=1.2\linewidth]{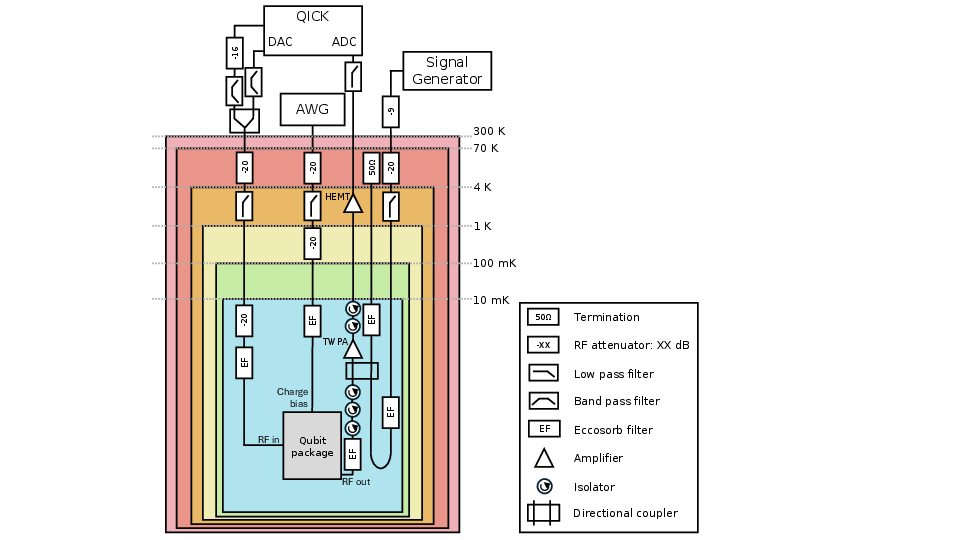}
\caption{Block diagram of experimental apparatus. The readout signals are carried on the QICK DAC line, with 16~dB of attenuation. The control pulses are carried on the other DAC line. There are four charge bias lines output from the AWG, one for each qubit. Each charge bias line is attenuated and filtered the same; only one line is shown in the diagram for visual clarity.}
\label{fig:RFdiag}
\end{figure*}

A diagram of the electronics and DR stages is shown in Fig.~\ref{fig:RFdiag}. 
The qubit package has one input and one output connection to a common transmission line that feeds all four resonators. 
Each qubit has a separate connection for its charge bias line. 
All input lines consist of stainless steel coaxial cables from room temperature to 10~mK with attenuation and IR filters \cite{spahn22} to limit thermal loading. 
Output lines are NbTi superconducting coaxial cable between the 10~mK and 4~K stages to minimize losses and stainless steel from 4~K to room temperature.
The output signals from the qubit package are amplified by a Traveling Wave Parametric Amplifier (TWPA) with a microwave pump tone at the 10~mK stage \cite{macklin15}, and by a high-electron-mobility transistor (HEMT) amplifier at the 4~K stage. 

Two different warm RF systems are used for qubit characterization and measurement. 
Both are connected to the qubit input and output ports via a warm RF switch (not shown in Figure \ref{fig:RFdiag}). 
A vector network analyzer (not shown in Fig. \ref{fig:RFdiag}) is used for continuous wave measurements during initial qubit characterization. Pulsed RF measurements are performed using the Quantum Instrumentation and Control Kit (QICK): a Xilinx RFSoC board with custom open-source software and firmware for the control of qubit systems \cite{qick_superconducting, qick}. The qubit charge bias voltage is supplied by an arbitrary waveform generator (AWG).
The system has since been upgraded to allow for multiplexed simultaneous readout, but at the time of the measurement presented in the main text, this was not possible. As such, individual qubits are measured consecutively for each bias voltage in a Ramsey tomography scan before moving onto the next voltage value. 

\section{The response of layered superconducting devices to ionizing events}
Consider the energies and time scales involved in the absorption and dissipation of ionizing radiation in a silicon substrate layered with superconducting aluminum.
Cosmic and gamma rays typically deposit hundreds of keV of energy in a chip. 
From the initial energy deposition, one electron-hole pair is generated per 3.7~eV deposited into the electronic system~\cite{ramanathan20} along with a burst of phonons as energy is transferred to the crystal lattice. 
The electrons and holes can recombine promptly or after some diffusion, producing more phonons. Alternatively, they can be trapped by impurities in the substrate material. 
If not collected, these trapped charges are quasi-stable, with lifetimes in the range of hours to days at mK temperatures~\cite{noah18}, and will alter the ambient electric field at the qubit island. 
The phonons, by comparison, become quasi-diffusive within 50~ns, then travel ballistically and potentially interact with the superconducting metal films making up the qubit resonator structures at the crystal surfaces.
At the superconducting film, these phonons are sufficiently energetic to break Cooper pairs, creating an excess population of Bogoliubov Quasiparticles (QPs). 
This excess QP population typically takes milliseconds to dissipate \cite{devissier11}.
Straightforward measurements of $T_1$, typically taking tens of microseconds, are therefore fast relative to both QP dissipation and timescales on which trapped charge is released. While the QP population can be inferred from $T_1$ measurements \cite{mcewen22}, the effects of charge movement and recombination in the substrate have a nontrivial time structure. 
Accordingly, the focus in this work is on observing fluctuations in offset charge over many hours. 

\section{Measurement Methodology}

To track charge bursts over time we use the same Ramsey tomography procedure as in Ref.~\cite{wilen21}: 
we apply a gate sequence of $X/2$--\textit{Idle}--$X/2$, followed by a readout pulse. 
During the idle time, the state vector acquires a phase that depends on $n_g$, as in Eq.~\ref{eq:phase}. 
We choose the idle time to be $1/(4\Delta f_{01})$, where $\Delta f_{01}$ is the maximum frequency separation of the two parity bands. 
The final $X/2$ pulse then maps the resulting state vector onto the $|1\rangle$ state and is probed via the readout resonator. 
The qubit is then passively initialized into the $|0\rangle$ state before the next measurement.

\begin{figure}[t!]
\centering
\includegraphics[width=\linewidth]{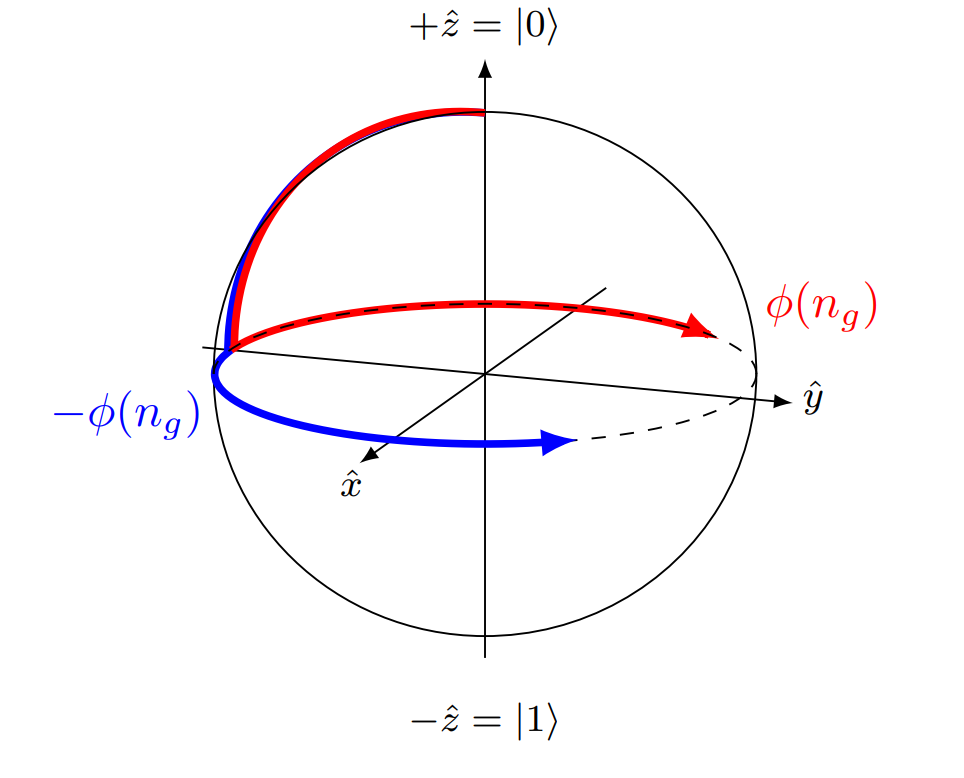}
\caption{Bloch sphere illustrating charge tomography sequence. Red and blue arrows indicate the parity-dependent path of the Bloch vector.}
\label{fig:bloch}
\end{figure}

This measurement sequence is averaged 200--300 times (depending on the qubit) for each applied offset charge value. 
At each applied bias voltage (corresponding to an unknown overall value of $n_g$), measurements are taken sequentially by qubit, meaning all 200 individual measurements are taken and averaged on qubit 1, then qubit 2, and so on, before the applied offset charge value is changed and the measurement cycle is repeated. 
Each individual Ramsey sequence takes 0.0049~s. Acquiring full statistics at each charge bias point takes 0.98--1.479~s per qubit. 
One full tomographic scan across all charge bias points takes 355~s.

\section{Jump-finding algorithm}
We utilize a data-driven method to identify charge jumps: a template for each qubit averages together 15--23 scans from the SC datasets that are free of charge jumps larger than 0.03$e$, the approximate detectable jump size given our specific averaging requirements.
The fluctuations for each selected jump-free set are randomly distributed in time. 
(Qubit 4 required two templates, one for SC and one for SO, due to reproducible noise differences between
shield configurations.) 
Templates are one period long, with a full scan created by stitching together period templates to the required length. The error for each point in the template is derived from the standard deviation across all template scans at that charge bias point. 

Once the template for each qubit is obtained, 
we find the best-fit phase and minimum reduced $\chi^2$ from fitting the template to the first $n$ points, for each value of $n$ in $(1,N)$:
\begin{equation}\label{eq:rollingchi}
    \chi^2_n(\theta) = \frac{1}{n} \sum_i^n \frac{(x_i - \hat{x}(\theta))^2}{\hat{\sigma}^2(\theta)}.
\end{equation}
The minimum value of $\chi^2_n(\theta)$ is calculated for each value of $n\in(1,N)$ where $N$ is the total number of points in the scan, with the value $\theta_{\rm min}^n$ corresponding to the phase associated with the minimum $\chi^2$ at that point. 

Charge jumps cause discontinuities in $\theta$. If a discontinuity is present, the quantity $\chi^2_n(\theta_{\rm min})$ will rapidly begin to increase with each subsequent term in the sum. 
When the rolling $\chi^2$ value exceeds a pre-set threshold (set individually for each qubit), a jump is identified and the procedure resets, with the first point over the threshold being reset as $i=1$ in Eq.\eqref{eq:rollingchi} and $N$ limited to the remaining number of points in the scan.
Because this is a cumulative process, it has limited efficiency in finding jumps in the first $\sim20$ points
of each scan.
The minimum jump size resolution is qubit-dependent since each qubit has a slightly different period length in $e$. This value varies from $0.026e$ to $0.029e$.


\begin{table}[t!]
\centering
\begin{tabular}{r | c}
& Efficiency \\\hline
Q1 & $0.83 \pm 0.01$ \\
Q2 & $0.79 \pm 0.01$ \\
Q3 & $0.87 \pm 0.03$ \\
Q4 (SC) & $0.72 \pm 0.02$ \\
Q4 (SO) & $0.74 \pm 0.05$ \\
\end{tabular}
\caption{Efficiencies of correctly identifying charge jumps with magnitude $0.1e \le |\Delta q| \le 0.5e$. Errors represent systematic variation in the ability to find jumps of different sizes in this range.  
}\label{tab:efficiency}
\end{table}

The efficiency of this method is assessed by applying it to synthetic data with charge jumps inserted at known locations.
For each qubit we simulate 1600 scans with a jump rate of 1.1~mHz (slightly higher than the actual rate to ensure we do not see effects of pile-up). The jumps are randomly injected with sizes selected from a flat probability distribution between 0.01$e$ and 0.5$e$.
The simulated scans are produced using the templates discussed above, convolved with a Gaussian noise spectrum to add the characteristic noise of our data. 
The accuracy of the jump-finding algorithm is gauged by the fraction of known charge jumps in the synthetic data that are correctly identified. 
Systematic errors on this quantity are derived from the standard deviation of this value for 15 different jump sizes across 75 sets of simulated scans per qubit (see Table~\ref{tab:efficiency}). 
This effectively captures the variation in efficiency across the range of $0.1e \le |\Delta q| \le 0.5e$, which can be crudely approximated by Gaussian (with slightly lower efficiency near the $0.1e$ threshold in particular). 

The $\chi^2$-threshold and $n$-delay discussed above are tuned to minimize false positives (e.g. noise tagged as jumps) while keeping the efficiency for tagging real jumps as high as possible. 
We calculate an efficiency for detecting real jumps of $>70\%$ for $|\Delta q|>0.1 e$. 
This efficiency is fundamentally limited by the difficulty of our method finding jumps at the start and end of individual scans; improvements will be the subject of future work.  
Event rates in this data are sufficiently low that systematic uncertainties in our efficiency determination are subdominant to our statistical uncertainties. 
Errors presented in the main text are therefore Poissonian.

\section{Radiation background characterization}

\begin{figure}
    \centering
    \includegraphics[width=\linewidth]{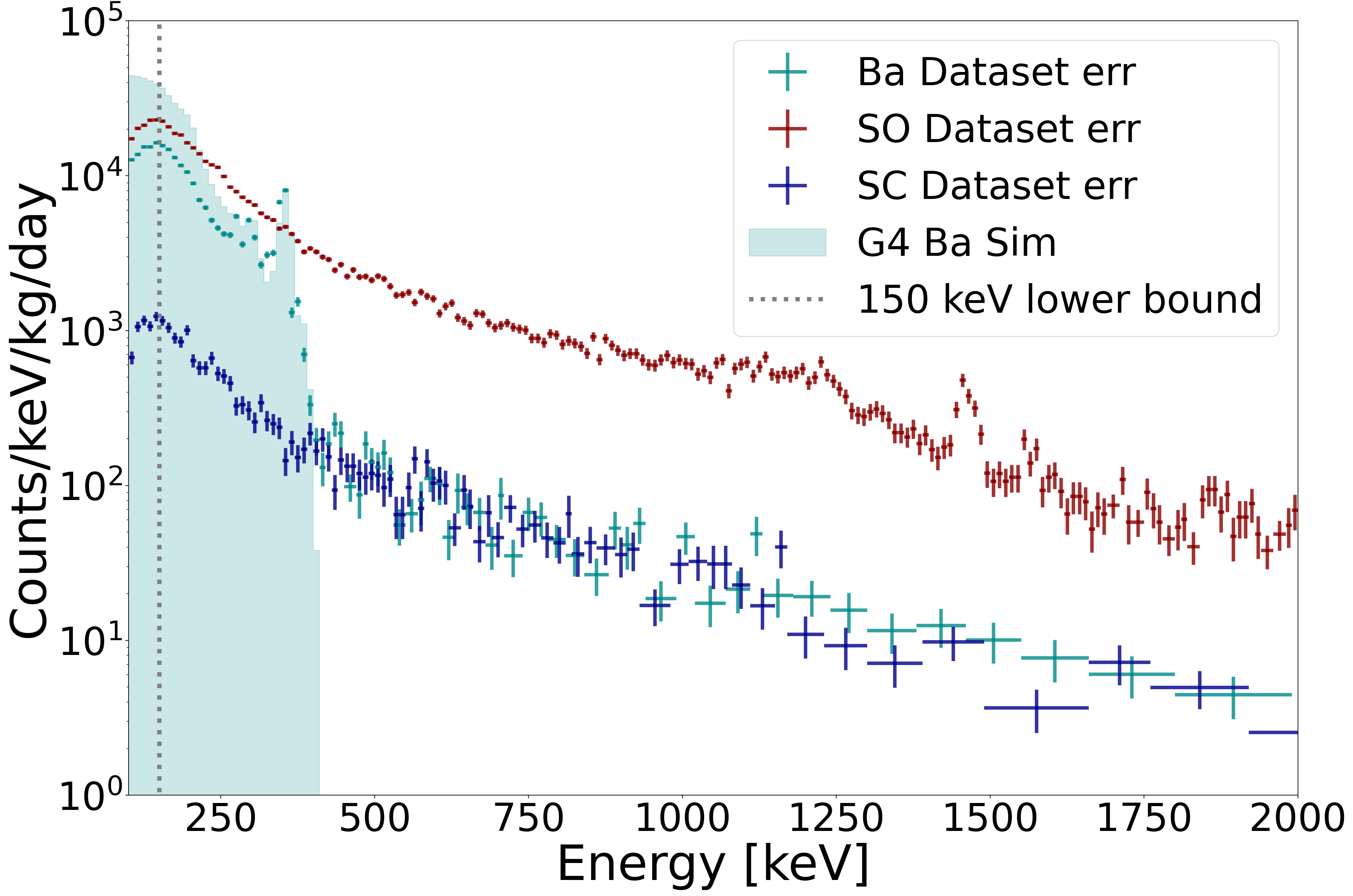}
    \caption{The energy spectra collected with a TES-coupled LMO detector are shown. Data collected in the SO configuration are shown in red, and SC configuration in blue. We acquired data with a $^{133}$Ba source (teal) to validate our analysis and compared it to a Geant4 simulation (teal, shaded). The vertical grey line at 150~keV indicates the threshold above which the response of the LMO detector is unitary and above which the ratio of events measured in the SO to SC dataset is found to be $20 \pm 1$.}
    \label{fig:LMO}
\end{figure}

We characterize the spectrum of radiation incident on the qubit package through the use of another
detector -- a Li$_2$MoO$_4$ (LMO) crystal read out with a Transition-Edge Sensor (TES)~\cite{LMO}. This detector is a cryogenic calorimetric device that uses a 2~cm cube of LMO with a mass of 21~g that has a 1~cm diameter, 400~nm thick gold film deposited on one of its sides through e-beam evaporation. A TES on a separate, $3\times 3\times 0.4$~mm Si chip is connected through a gold wire bond and measures the energy depositions from gammas in the LMO through measurement of its temperature~\cite{Chen:2023,Doug:2023}. The LMO and the qubit chip operate simultaneously inside the same DR. Thus, the LMO data provides a direct measurement of the gamma flux 18.7~cm away from the qubit chip. 
The energy spectrum for SC and SO data, as recorded by the LMO device, are shown in Figure~\ref{fig:LMO}.
The integral of these spectra above 150~keV (the low-energy efficiency of the LMO detector is not yet fully characterized) is used to calculate the gamma flux ratio between SO to SC. This ratio is insensitive to systematic errors related to energy-independent detector response, acceptance, and live-time effects and represents a direct measurement of the effect of the NEXUS shield on radiation flux.
By closing the lead shield, the overall gamma flux is reduced in the LMO crystal by a factor of $A_{\rm LMO} = 20\pm1$ for energies above 150~keV. 
The energy scale in the SO data is calibrated using the $^{40}$K peak at 1460~keV. 

To calibrate the SC data,  which does not have any peaks to use for calibration, we include data taken with a $^{133}$Ba source inside the lead shield and compare with a \texttt{GEANT4} \cite{AGOSTINELLI2003250} simulation. The two $^{133}$Ba and SC datasets were taken with the LMO device in the same bias conditions. The 356~keV peak is used to set the energy scale in the $^{133}$Ba data, and the same energy calibration is used for the SC dataset. The \texttt{GEANT4} simulation only simulates the $^{133}$Ba source, not the ambient SC spectrum. 
As shown in Figure~\ref{fig:LMO}, the Ba dataset agrees very well with the SC dataset above 400~keV; above that rough threshold there is no flux from the $^{133}$Ba source.

At lower energies, the \texttt{GEANT4} simulation predicts a higher flux than we see in the LMO, which is likely a combination of the non-unity efficiency of the LMO at lower energies and an insufficiently detailed model in \texttt{GEANT4} of the material interposed between the $^{133}$Ba source and the LMO. As mentioned above, by taking the ratio of SO and SC data we make a flux ratio which is insensitive to most of these systematic effects.

We exploit the measured ratio between the SO and SC LMO datasets $A_{\rm LMO}$ by solving for a constant excess charge jump rate $R_{\rm excess}$ present in both datasets and the gamma-induced charge jump rate for each. By solving, 
\begin{equation}
\begin{split}
    R^{SO} &= R^{SO}_{\gamma} + R_{\rm excess} \\ 
    R^{SC} &= R^{SC}_{\gamma} + R_{\rm excess} \\ 
    R^{SO} &= A_{\rm LMO} \times R^{SC}_{\gamma},
\end{split}
\end{equation}
where $R^{SO/SC}$ is the measured charge jump rate in the SO and SC data and $R^{SO/SC}_{\gamma}$ is corrected gamma-induced rate in each dataset, we find an excess charge jump rate of $R_{\rm excess} = 0.17^{+0.04}_{-0.03}$. 
These results are presented in Table~\ref{tab:sing_rates} of the main text.

\end{document}